\documentclass[12pt,a4paper]{article}
\usepackage{graphicx}
\usepackage{epsfig}

\begin{document}

\begin{center}
 {\bf \Large $NN$ potentials from inverse scattering in the $J$-matrix
 approach. }\\[3pt]
{\large  S.~A.~Zaitsev, E.~I.~Kramar}\\[2pt]
{\sl Department of Physics, Khabarovsk State Technical University,\\
Tikhookeanskaya 136, Khabarovsk 680035, Russia}\\[4pt]

\end{center}


\begin{abstract}
An approximate inverse scattering method \cite{TMP1,TMP2} has been used to
construct separable potentials with the Laguerre form factors. As an
application, we invert the phase shifts of proton-proton in the $^1S_0$ and
$^3P_2-^3F_2$ channels and neutron-proton in the $^3S_1-^3D_1$ channel elastic
scattering. In the latter case the deuteron wave function of a realistic $np$
potential was used as input.
\end{abstract}

\section{Introduction}
Low-rank separable potentials greatly simplify many-body computations. The main
sources of separable nucleon-nucleon ($NN$) potentials are the inverse
scattering methods (see \cite{SepNN} and references therein). On the other
hand the model effective two-particles interactions (see e.g. \cite{Shell})
constructed with use of a realistic $NN$ potential (e.g. \cite{Nijmegen} ) as
input are also of separable form.

\par Recently a novel method was proposed for the treating of three-body
problems including Coulomb forces \cite{TB}. In the framework of this approach
the potential operator of the short-range part of full interaction is expanded
with use of Coulomb-Sturmian (Laguerre) functions. Whereas Coulomb interaction
is kept in the Green's operator involved in Lippman-Schwinger equation.
Instead of separable potentials (of relatively high rank) resulting from a
realistic $NN$ interaction, the potentials obtained by means of inverse
scattering technique \cite{TMP1,TMP2} within so-called $J$-matrix method can
be applied in the three-body calculations \cite{TB}.

\par The $J$-matrix method \cite{Jmtx,AVMRG,HOR} is equivalent to the potential
separable expansion approach \cite{PSE,Papp}; here too the potential
$\widehat{V}$ is approximated by its projection $\widehat{V}^N$ onto the
finite subspace spanned by the first $N$ (Laguerre or oscillator ) basis
functions. However in the $J$-matrix method the Schr\"{o}dinger equation
rather than Lippman-Schwinger one is solved in the square integrable basis
representation. In the $J$-matrix approach an auxiliary function have been
introduced resulting from discrete representation of the full Green's operator
\cite{BR}. This function (or matrix in the multichannel scattering case),
so-called ${\cal P}$-matrix, plays in our inversion scheme a part similar to
that of the $R$-matrix in the potential reconstruction in the framework of
finite-difference analogue of $R$-matrix scattering theory \cite{Zakhariev}.
In particular, information requisite for the potential construction can also
be obtained from the poles and residues of the ${\cal P}$-matrix.

\par As an application of our method, we invert phase shifts of the
proton-proton and neutron-proton elastic scattering. In the $NN$ potential
construction an attempt have been made to account for the deuteron wave
function \cite{Nijmegen} using the phase-equivalent transformation
\cite{PHT,TMP2}.

\par In Sect. 2 we outline the $J$-matrix formalism for the solution of
multichannel scattering problem. In Sect. 3 we briefly present the inversion
procedure in the Laguerre form factors case, and describe the taking account
for deuteron wave function technique within our method. We then apply the
method to the $NN$ elastic scattering in Sect. 4. In Sect. 5 we summarize our
conclusions.

\section{Elements of the $J$-matrix formalism}
The formalism is spelled out in detail in \cite{BR} for treating the problem
of multichannel scattering from the potentials of the form
\begin{equation}
\widehat{V}= \sum_{\alpha, \: \beta \,= \, 1,\,2} \left| \alpha \right>
V^{\alpha \: \beta} \left< \beta \right|, \label{GPot}
\end{equation}
where the partial wave potentials $V^{\alpha \: \beta}$ are given by the
expansion
\begin{equation}
V^{\alpha \:  \beta}=\frac{\hbar^2}{2\mu}
\sum_{n=0}^{N_{\alpha}-1}\sum_{n'=0}^{N_{\beta}-1} \left|
\overline{\phi}_n^{(\alpha)}\right> V_{n, \, n'}^{\alpha \: \beta} \left<
\overline{\phi}_{n'}^{(\beta)}\right|. \label{PPot}
\end{equation}
Here, the $\alpha$ and $\beta$ denote sets of quantum numbers including the
values of the orbital momenta $l_{\alpha}$ and $l_{\beta}$, and
\begin{equation}
  \left| \overline{\phi}_n^{(\alpha)} \right>= \frac{\displaystyle n!}
  {\displaystyle r \, (n+2l_{\alpha}+1)!} (a_{\alpha}  r)^{l_{\alpha}+1} \,
  e^{-a_{\alpha}  r/ 2} L_n^{2l_{\alpha}+1}(a_{\alpha} r), \label{orthbf}
\end{equation}
are the functions bi-orthogonal to the Laguerre basis functions $\left|
\phi_n^{(\alpha)} \right>$ \cite{Jmtx} in the channel $\alpha$
\begin{equation}
  \left| \phi_n^{(\alpha)} \right> = (a_{\alpha}  r)^{l_{\alpha}+1} \,
  e^{-a_{\alpha}  r/ 2} L_n^{2l_{\alpha}+1}(a_{\alpha} r), \label{basfun}
\end{equation}
i.e. $\left<\phi_n^{(\alpha)} \left| \overline{\phi}_{n'}^{(\alpha)} \right.
\right>= \delta_{n \, n'}$. Here, $a_{\alpha}$ is a scaling parameter, $\mu$
is the reduced mass of the system. We confine ourselves to the elastic
scattering case, in which the energy $\epsilon \equiv k^2 =$ $2\mu/\hbar^2\, E$
is the same in the two channels.

\par The elements $u_{\alpha \beta}(k, \, r)$ of the $2 \times 2$ matrix as
the solution of the Schr\"{o}dinger equation are represented in the $J$-matrix
method in the form of the expansion
\begin{equation}
u_{\alpha \, \beta}(k,r) = \sum^{\infty}_{n=0} \left| \phi_n^{(\alpha)}
\right> b^{(\alpha \, \beta)}_n(k), \label{fser}
\end{equation}
whose coefficients $b^{(\alpha \, \beta)}_n$ obey the relations \cite{BR}:
\begin{equation}
  b^{(\alpha\beta)}_n(k)= \left\{
    \begin{array}{lr}
\sum \limits _{\alpha'} P^{\alpha \, \alpha'}_{n, \: N_{\alpha'}-1} (\epsilon)
\, J_{N_{\alpha'}-1, \, N_{\alpha'}}^{(\alpha')}(k) \, b^{(\alpha' \,
\beta)}_{N_{\alpha'}}(k),
&  n \le N_{\alpha}-1, \\[4mm]
S_n^{(\alpha)}(k) \delta_{\alpha \beta} + C_n^{(\alpha)}(k) \, K_{\alpha \,
\beta}, &   n \ge N_{\alpha}-1.
     \end{array}
             \right. \label{bsol}
\end{equation}
Here, $K_{\alpha \, \beta}$ are the elements of the $K$-matrix, $P^{\alpha \,
\beta}_{n,\, n'}$ are the elements of the matrix
\begin{equation}
{\bf P}= \left( {\bf R} \epsilon - {\bf h} \right)^{-1}, \label{Pel}
\end{equation}
where ${\bf h}$ is the matrix of the Hamiltonian $\widehat{h} =
\widehat{h}_0^{(1)}+\widehat{h}_0^{(2)}+ \frac{\displaystyle 2 \,
\mu}{\displaystyle \hbar^2}\widehat{V}$ calculated in the basis $\left\{
\left| \alpha \right> \left| \phi_n^{(\alpha)} \right>\right.$,
$n=\overline{0, \: N_{\alpha}-1}$, $ \left. \alpha = \overline{1, \: 2}
\right\}$. ${\bf R}$ is the block-diagonal overlap matrix involving the
symmetric tridiagonal submatrices ${\bf R}^{(\alpha)}$ \cite{Broad}:
\begin{equation}
 \begin{array}{c}
  R_{n, \, n}^{(\alpha)}= \frac{\displaystyle 2}{\displaystyle
  a_{\alpha}}(n+l_{\alpha}+1)
  \frac{\displaystyle (n+2l_{\alpha}+1)!}{\displaystyle n!}, \\[3mm]
  R_{n, \, n+1}^{(\alpha)}=R_{n+1, \, n}^{(\alpha)}= -\frac{\displaystyle
  1}{\displaystyle a_{\alpha}}
   \frac{\displaystyle (n+2l_{\alpha}+2)!}{\displaystyle n!}.\\
 \end{array}
\label{Rij}
\end{equation}
The matrix ${\bf h}_0^{(\alpha)}$ of free Coulomb Hamiltonian
\begin{equation}
\widehat{h}_0^{(\alpha)} = -\frac{\displaystyle d^2}{\displaystyle d\,r^2}+
\frac{\displaystyle l_{\alpha}(l_{\alpha}+1)}{\displaystyle r^2} +
\frac{\displaystyle 2 \, \tau \, k}{\displaystyle r}, \label{h0}
\end{equation}
where $\tau=\frac{\displaystyle Z\, e^2 \mu}{\displaystyle \hbar^2 \, k}$, is
also symmetric and tridiagonal \cite{Broad}:
\begin{equation}
  \begin{array}{c}
   \left[ {\bf h}_0^{(\alpha)} \right]_{n, \, n}= \frac{\displaystyle
   a_{\alpha}^2}{\displaystyle 4}R_{n, \, n}^{(\alpha)}
  +2\tau k \frac{\displaystyle (n+2l_{\alpha}+1)!}{\displaystyle n!},\\[3mm]
 \left[ {\bf h}_0^{(\alpha)} \right]_{n, \, n+1}= \left[ {\bf h}_0^{(\alpha)}
 \right]_{n+1, \, n}= -\frac{\displaystyle a_{\alpha}^2}
 {\displaystyle 4}R_{n, \, n+1}^{(\alpha)}.\\
 \end{array}
\label{h0ij}
\end{equation}
$J_{N_{\alpha}-1, \, N_{\alpha}}^{(\alpha)}$ are the elements of so called
$J-$matrix \cite{Jmtx}:
\begin{equation}
  J_{n, \, n'}^{(\alpha)}(k)= \left[ {\bf h}_0^{(\alpha)} \right]_{n, \, n'}-
  k^2R_{n, \, n'}^{(\alpha)}. \label{Jme}
\end{equation}

\par In what follows, ${\bf h}$ denotes an
${\cal N} \times {\cal N}$-matrix (${\cal N}= N_1+N_2$) whose elements
$h_{ij}$ are defined in accordance with the rule $h_{ij}=h^{\alpha \, \beta}
_{n \, n'}$, if $i= n+\sum \limits _{\alpha' < \alpha}N_{\alpha'}+1$ and $j=
n'+\sum \limits _{\alpha' < \beta}N_{\alpha'}+1$.

\par With the spectral decomposition of the matrix
${\bf h}'={\bf Z}{\bf \Lambda}{\bf Z}^*$ (where superscript $*$ denotes
transposition) of the Hamiltonian $\widehat{h}$ calculated in the
orthonormalized basis $\left\{ \left| \alpha \right> \left| \psi_n^{(\alpha)}
\right> \right.$, $n=$ $\overline{0, \: N_{\alpha}-1}$, $\left. \alpha =
\overline{1, \: 2} \right\}$, where ${\bf Z}$ is the orthogonal matrix of
eigenvectors, and ${\bf \Lambda}$ is the diagonal matrix of the eigenvalues
$\left\{ \lambda_j \right\}$ of the matrix ${\bf h}'$, expression (\ref{Pel})
can be written as \cite{Broad}
\begin{equation}
{\bf P}={\bf
D}^*{\bf Z} \left[ {\bf I} \epsilon- {\bf \Lambda}\right]^{-1} {\bf Z}^*{\bf
D}.
\label{Pm}
\end{equation}
Here, ${\bf I}$ is the identity matrix, ${\bf D}$ is the block-diagonal matrix
involving the $N_{\alpha} \times N_{\alpha}$ submatrices ${\bf D}^{(\alpha)}$,
$\alpha=1,\,2$ of the transformation to the orthonormal basis $\left\{
\left|\psi_n^{(\alpha)} \right> \right.$, $\left. n=\overline{0, \,
N_{\alpha}-1} \right\}$, i. e.
\begin{equation}
\left|\psi_n^{(\alpha)} \right>=\sum \limits_{j=1} ^{N_{\alpha}} D_{n+1, \,
j}^{(\alpha)} \left|\phi_{j-1}^{(\alpha)}\right>. \label{ON}
\end{equation}

\par $C_n^{(\alpha)}$ and $S_n^{(\alpha)}$ coincide respectively with
the real and image part of the solution $C_n^{(\alpha)(+)}$ of the discrete
analogue of the Coulomb Schr\"{o}odinger equation \cite{Jmtx}
\begin{equation}
 \begin{array}{l}
C_n^{(\alpha)(+)}(k) = - \frac{ \displaystyle n! \, e^{ \mbox{i}
\sigma_{\alpha} } \, e^{(\theta_{\alpha}-\pi/2)\tau} \, e^{-\mbox{i}(n+1) \,
\theta_{\alpha} } } { \displaystyle \Gamma(n+l_{\alpha}+2+\mbox{i} \tau)
\left(2\, \sin \theta_{\alpha} \right)^{l_{\alpha}}
} \\
\qquad \qquad \qquad \qquad \qquad \qquad \qquad \qquad \times
{_2F_1}(-l_{\alpha}+\mbox{i}\tau, \, n+1; \, n+l_{\alpha}+2+\mbox{i}\tau; \,
e^{-2\mbox{i}\theta_{\alpha} }). \\
 \end{array}
\label{Cp}
\end{equation}
Here, $\sigma_{\alpha}=arg \Gamma(\mbox{i}\tau+l_{\alpha}+1)$, $e^{\mbox{i}
\theta_{\alpha} }=(\mbox{i}q_{\alpha}-\frac12)/ (\mbox{i}q_{\alpha}+\frac12)$,
$q_{\alpha}=k/a_{\alpha}$. For $Z=0$ hypergeometric function $_2F_1$ in Eq.
(\ref{Cp}) degenerates into a polynomial of degree $l_{\alpha}$ in
$e^{-2\mbox{i}\theta_{\alpha} }$.

\par The symmetric $2 \times 2$ $K$-matrix corresponding to the potential
(\ref{GPot}), (\ref{PPot}) has the form \cite{Broad}:
\begin{equation}
{\bf K}=- \left[ {\bf C}_{N-1} - {\bf {\cal P}}{\bf J}{\bf C}_N \right]^{-1}
\left[ {\bf S}_{N-1} - {\bf {\cal P}} {\bf J}{\bf S}_N \right], \label{Kmtx}
\end{equation}
where
\begin{equation}
 \begin{array}{c}
\left[ {\bf C}_{N-1} - {\bf {\cal P}}{\bf J}{\bf C}_N \right]_ {\alpha, \,
\beta} = C^{(\alpha)}_{N_{\alpha}-1}(k) \, \delta_{\alpha, \, \beta} -{\cal
P}_{\alpha, \, \beta}(\epsilon) \, J_{N_{\beta}-1, \, N_{\beta}}^{(\beta)}(k)
\,
C^{(\beta)}_{N_{\beta}}(k), \\[3mm]
\left[ {\bf S}_{N-1} - {\bf {\cal P}}{\bf J}{\bf S}_N \right]_ {\alpha, \,
\beta} = S^{(\alpha)}_{N_{\alpha}-1}(k) \, \delta_{\alpha, \, \beta} -{\cal
P}_{\alpha, \, \beta}(\epsilon) \, J_{N_{\beta}-1, \, N_{\beta}}^{(\beta)}(k)
\,
S^{(\beta)}_{N_{\beta}}(k), \\
 \end{array}
\end{equation}
${\cal P}_{\alpha, \, \beta}(\epsilon) \equiv P^{\alpha \, \beta}_
{N_{\alpha}-1,\, N_{\beta}-1}(\epsilon)$ are the elements of so called ${\cal
P}$-matrix.

\section{Description of the method}
\par By choosing the orthonormal basis functions
$\left\{ \left|\psi_n^{(\alpha)} \right> \right\}$ (\ref{ON}) in the form
\begin{equation}
\left|\psi_n^{(\alpha)} \right>= \left( \frac{\displaystyle a_{\alpha} \, n!}
{\displaystyle (n+2l_{\alpha}+2)!} \right)^{1/2} \,
(a_{\alpha}r)^{l_{\alpha}+1}\, e^{-a_{\alpha}r/2}
L_n^{2l_{\alpha}+2}(a_{\alpha}r), \label{bf1}
\end{equation}
which is to say that the matrices ${\bf D}^{(\alpha)}$ are defined by the
expressions
\begin{equation}
D_{i, \, j}^{(\alpha)}= \left\{
 \begin{array}{lr}
  d_{i-1}^{(\alpha)}, & i \ge j, \\
  0, & i < j,\\
 \end{array}
\right. \qquad
\end{equation}
where
\begin{equation}
d_n^{(\alpha)}= \left( \frac{\displaystyle a_{\alpha} \, n!} {\displaystyle
(n+2l_{\alpha}+2)!} \right)^{1/2}, \label{dn}
\end{equation}
we derive the following expressions for the ${\cal P}$-matrix elements
\begin{equation}
 \begin{array}{c}
{\cal P}_{11}(\epsilon)= \left( d_{N_1-1}^{(1)} \right)^2 \, \sum \limits
_{j=1}^{ {\cal N } } \frac{\displaystyle Z_{N_1, \: j}^{\, 2}}
{\displaystyle \epsilon -  \lambda_j}, \\[4mm]
{\cal P}_{22}(\epsilon)= \left( d_{N_2-1}^{(2)} \right)^2 \, \sum \limits
_{j=1}^{ {\cal N } } \frac{\displaystyle Z_{{\cal N}, \: j}^{\,
2}}{\displaystyle \epsilon -
 \lambda_j},\\[4mm]
{\cal P}_{12}(\epsilon)= {\cal P}_{21}(\epsilon) = d_{N_1-1}^{(1)} \,
d_{N_2-1}^{(2)}  \, \sum \limits _{j=1} ^{ {\cal N} } \frac{\displaystyle
Z_{N_1, \: j} \, Z_{{\cal N}, \: j}}
{\displaystyle \epsilon - \lambda_j}.\\
 \end{array}
\label{Pab}
\end{equation}

\par It should be noted that the elements ${\cal P}_{\alpha, \,
\beta}(\epsilon)$ (\ref{Pab}) decay ($\sim \epsilon^{-1}$) at the energies
$\epsilon
> \lambda_{{\cal N}}$, and the same is true for the $K$-matrix elements
(\ref{Kmtx}). On this basis, we set ourselves the task of the construction of
the potential (\ref{GPot}), (\ref{PPot}) descriptive of the $K$-matrix in an
energy interval $[0, \, \epsilon_0]$ and the system bound spectrum. In the case
of arbitrary $K$-matrix behaviour it is suggested that $\epsilon_0 <
\lambda_{\cal N}$.

\par The expressions (\ref{Pab}) for  the ${\cal P}$-matrix are similar in
structure to those for the $R$-matrix \cite{Rmtx}. This analogue have inspired
us \cite{TMP2} to implement the technique used within the finite-difference
analogue of the $R$-matrix scattering theory \cite{Zakhariev} for the
retrieval of information requisite for the potential constructing from the
$K$-matrix. For this purpose we fix some values of $a_{\alpha}, \, N_{\alpha}$
and define functions
\begin{equation}
 \begin{array}{c}
\widetilde {\cal P}_{11}(\epsilon) = \frac{\displaystyle 1}{\displaystyle
\Delta(k) \, J_{N_1-1, \, N_1}^{(1)}}(k) \left \{ \left(S_{N_{1}-1}^{(1)}(k)+
C_{N_{1}-1}^{(1)}(k) K_{11} \right) \left(S_{N_{2}}^{(2)}(k)+
C_{N_{2}}^{(2)}(k)K_{22} \right) \right. \\
\phantom{CCCCCCCCCCCCCCCCCCCCCCCCCCCCCCCCCC} \left.
-C_{N_{1}-1}^{(1)}(k)C_{N_{2}}^{(2)}(k)K^2_{12} \right\},\\ \end{array}
\label{P11t}
\end{equation}
\begin{equation}
 \begin{array}{c}
\widetilde {\cal P}_{22}(\epsilon) = \frac{\displaystyle 1}{\displaystyle
\Delta(k) \, J_{N_2-1, \, N_2}^{(2)}}(k) \left \{ \left(S_{N_{2}-1}^{(2)}(k)+
C_{N_{2}-1}^{(2)}(k) K_{22} \right) \left(S_{N_{1}}^{(1)}(k)+
C_{N_{1}}^{(1)}(k)K_{11} \right) \right.\\
\phantom{CCCCCCCCCCCCCCCCCCCCCCCCCCCCCCCCCC}
\left.-C_{N_{2}-1}^{(2)}(k)C_{N_{1}}^{(1)}(k)K^2_{12} \right\}, \\
 \end{array}
\label{P22t}
\end{equation}
\begin{equation}
\widetilde {\cal P}_{12}(\epsilon)=\widetilde {\cal P}_{21}(\epsilon)
=-\frac{\displaystyle k \, K_{12} } {\displaystyle \Delta(k) \, J_{N_1-1, \,
N_1}^{(1)}(k) \, J_{N_2-1, \, N_2}^{(2)}(k) }, \label{P12t}
\end{equation}
where
\begin{equation}
\Delta(k)= \left(S_{N_{1}}^{(1)}(k)+ C_{N_{1}}^{(1)}(k) K_{11} \right)
\left(S_{N_{2}}^{(2)}(k)+ C_{N_{2}}^{(2)}(k)K_{22} \right) -
C_{N_{1}}^{(1)}(k)C_{N_{2}}^{(2)}(k)K^2_{12}, \label{D}
\end{equation}
by inverting expression (\ref{Kmtx}) relative to the ${\cal P}$-matrix
elements. We set, in view of Eq. (\ref{Pab}),
\begin{equation}
 \begin{array}{c}
\lambda_j=\widetilde{\lambda}_j, \; Z_{N_1,\, j} = \left( d_{N_1-1}^{(1)}
\right) ^{-1} \sqrt{\mathop{Res} \limits _{\epsilon= \tilde \lambda _j}
\widetilde{\cal P}_{11} }(\epsilon),\\[3mm]
Z_{{\cal N},\, j} = \left( d_{N_2-1}^{(2)} \right) ^{-1} \left. \mathop{Res}
\limits _{\epsilon= \tilde \lambda _j} \widetilde{\cal P}_{12}(\epsilon)
\right/ \sqrt{\mathop{Res} \limits _{\epsilon= \tilde \lambda _j}
\widetilde{\cal P}_{11}(\epsilon) }, \;
j=\overline{1,\, p}, \\
 \end{array}
\label{inPar}
\end{equation}
where $\left\{ \widetilde{\lambda}_j < \epsilon_0, \, j=\overline{1, \,p}
\right\}$ are the ``internal'' poles of the functions $\widetilde{\cal
P}_{\alpha, \, \beta}$ (\ref{P11t})-(\ref{D}).

\par An search for optimal combinations $a_{\alpha}, \, N_{\alpha}$ is an important
constituent of the method. Assuming that the absolute values of the
off-diagonal elements of the $K$-matrix are negligibly small, this problem can
be solved in every channel $\alpha=1, \, 2$ separately. For this purpose, on
every step of variational procedure involving \cite{TMP1} in which the
rational approximant
\begin{equation}
 \begin{array}{c}
  {\cal P}_{\alpha}(\epsilon)= \left( d_{N_{\alpha}-1}^{(\alpha)} \right)^2 \,
  \sum \limits _{j=1}^{ N_{\alpha}} \frac{\displaystyle \left( Z_{N_{\alpha},
  \:
j}^{(\alpha)}\right)^{\, 2}} {\displaystyle \epsilon -
\lambda_j^{(\alpha)}},\\[3mm]
 \sum \limits _{j=1}^{N_{\alpha}} \left( Z_{N_{\alpha}, \, j} ^{(\alpha)}
 \right)^2 =1\\
 \end{array}
  \label{Pa}
\end{equation}
on the interval $[0, \, \epsilon_0]$ of the function
\begin{equation}
 \begin{array}{c}
\widetilde {\cal P}_{\alpha}(\epsilon)= \frac{\displaystyle 1} {\displaystyle
J_{N_{\alpha}-1, \, N_{\alpha}}^{(\alpha)}(k)} \frac{\displaystyle
\widetilde{C}_{N_{\alpha}-1}^{(\alpha)}(k) }
{\displaystyle \widetilde{C}_{N_{\alpha}}^{(\alpha)}(k) }, \\[5mm]
\widetilde{C}_n^{(\alpha)}(k) = \cos \delta_{\alpha} \, S_n^{(\alpha)}(k)+
\sin \delta_{\alpha} \, C_n^{(\alpha)}(k)\\
 \end{array}
\label{Pta}
\end{equation}
is constructed. Going to the next step, we increase $a_{\alpha}$ and decrease
$N_{\alpha}$. The condition that remainder of the approximation \cite{Sk} be
small is criteria for going to new combination $a_{\alpha}, \, N_{\alpha}$.

\par $N_{\alpha}$ is bounded below by the requirement that the
residues of $\widetilde{\cal P}_{\alpha}$ be positive, or by the equivalent
condition that the roots of the denominators of the functions $\widetilde{\cal
P}_{\alpha}$ alternate those of the numerators \cite{TMP1}. It is evident from
asymptotic expressions for the solutions $S_n^{(\alpha)}, \, C_n^{(\alpha)}$
as $n \rightarrow \infty$ ($k$ is bounded) \cite{Broad}:
\begin{equation}
 \begin{array}{c}
S_n^{(\alpha)}(k)=-n^{-(l_{\alpha}+1)} \sin \left\{ \sigma_{\alpha}
-(n+l_{\alpha}+1) \theta_{\alpha} - \tau \ln(2\, n \sin \theta_{\alpha}) +
\frac{\displaystyle \pi \, l_{\alpha}}{\displaystyle 2} \right \}, \\[3mm]
C_n^{(\alpha)}(k)=-n^{-(l_{\alpha}+1)} \cos \left\{ \sigma_{\alpha}
-(n+l_{\alpha}+1) \theta_{\alpha} -\tau \ln(2\, n \sin \theta_{\alpha}) +
\frac{\displaystyle \pi \, l_{\alpha}}{\displaystyle 2} \right \}, \\[3mm]
\widetilde{C} _n^{(\alpha)}(k)=-n^{-(l_{\alpha}+1)} \sin \left\{
\sigma_{\alpha} -(n+l_{\alpha}+1) \theta_{\alpha} - \tau \ln(2\, n \sin
\theta_{\alpha}) + \frac{\displaystyle \pi \, l_{\alpha}}{\displaystyle 2}+
\delta_{\alpha}  \right \}, \\
 \end{array}
\label{Asimp}
\end{equation}
that if $N_{\alpha}$ is large enough to fulfill the condition
\begin{equation}
\left| \frac{\displaystyle d}{\displaystyle d\,k} \left\{ \sigma_{\alpha}
-(N_{\alpha}+l_{\alpha}) \theta_{\alpha} -\tau \ln[2\, (N_{\alpha}-1) \sin
\theta_{\alpha}] \right\} \right| > \left|\frac{\displaystyle d \,
\delta_{\alpha} }{\displaystyle d\,k} \right|, \label{SepCond}
\end{equation}
the roots separation holds.

\par The parameters $a_{\alpha}, \, N_{\alpha}$ obtained above are
used in the coupled-channel case. Once the ``internal'' parameters has been
fixed (\ref{inPar}), $r={\cal N}-p$ ``external''  triplets $\left\{ \lambda_j
> \epsilon_0 \right.$, $Z_{N_1, \, j}$, $\left.Z_{{\cal N}, \, j}, \;
j=\overline{p+1, \, {\cal N}} \right\}$ remain unknown. Since the rows
$\left\{Z_{N_1, \, j} \right\}$ and $\left\{Z_{{\cal N}, \, j} \right\}$ of
the orthogonal matrix ${\bf Z}$ must meet the orthonormalization conditions
\begin{equation}
\sum \limits _{j=1}^{{\cal N}} Z^2_{N_1, \: j}= \sum \limits _{j=1}^{{\cal N}}
Z^2_{{\cal N}, \: j}=1, \quad \sum \limits _{j=1}^{{\cal N}} Z_{N_1, \: j} \:
Z_{{\cal N}, \: j}=0, \label{Norm}
\end{equation}
it is evident that the minimal r is equal to 2.
\par The values of the ``external'' parameters $\left\{ \lambda_j
\right\}$, as a rule, lie in the energy range that is far removed from the
right limit of the interval $[0, \, \epsilon_0]$. If the assumption is made
that for this energies the phase shifts and the mixing parameter are
negligible, for the parameters $\left\{ \lambda_j  > \epsilon_0 \right\}$ the
``external'' parameters $\left\{ \lambda_j^{(\alpha)} \right\}$ obtained in
the construction of the rank $N_{\alpha}$ potential in every channel
$\alpha=1, \, 2$ can be used.

\par There is for every bound state with the energy
$\varepsilon_{\nu}= -\kappa^2_{\nu}$ a corresponding parameter
$\lambda_{\nu}$: $\varepsilon_{\nu} < \lambda_{\nu} < \lambda_1$, which fits
the equation \cite{TMP3}:
\begin{equation}
\det \left[ {\bf C}_{N-1}^{(+)} - {\bf {\cal P}}{\bf J}{\bf C}_N^{(+)} \right]
= 0, \label{Bcond}
\end{equation}
where
\begin{equation}
\left[ {\bf C}_{N-1}^{(+)} - {\bf {\cal P}}{\bf J}{\bf C}_N^{(+)} \right]_
{\alpha, \, \beta} = C^{(\alpha)(+)}_{N_{\alpha}-1}(\mbox{i}\kappa_{\nu}) \,
\delta_{\alpha, \, \beta} -{\cal P}_{\alpha, \, \beta}(\varepsilon_{\nu}) \,
J_{N_{\beta}-1, \, N_{\beta}}^{(\beta)}(\mbox{i}\kappa_{\nu}) \,
C^{(\beta)(+)}_{N_{\beta}}(\mbox{i}\kappa_{\nu}).
\end{equation}
Eq. (\ref{Bcond}) is similar to the equation \cite{PSE,Papp}
\begin{equation}
\det [{\bf I}- {\bf G}^{(0)(+)}(\mbox{i}\kappa_{\nu}){\bf V}]=0, \label{Fred}
\end{equation}
where
\begin{equation}
[{\bf G}^{(0)(+)}(\mbox{i}\kappa_{\nu})]^{\alpha, \, \beta}_{n, \, n'}=
\delta_{\alpha, \, \beta}\langle \overline{\phi}_n^{(\alpha)} \left|
\widehat{G}^{(0)(+)}(\mbox{i}\kappa_{\nu}) \right|
\overline{\phi}_{n'}^{(\alpha)}\rangle
\end{equation}
are the matrix elements of the Coulomb Green's operator. In spite of the fact
that Eq. (\ref{Bcond}) possesses the redundant roots $\kappa=a_{\alpha}/2$ of
multiplicity $N_{\alpha}$, it is preferable to Eq. (\ref{Fred}), since the
desired potential matrix ${\bf V}$ not enters explicitly into Eq.
(\ref{Bcond}).

\subsection*{Phase-equivalent transformation}

From Eq. (\ref{Pab}) follows the expression that relates the matrices of the
phase-equivalent potentials (for comparison see \cite{PHT,TMP2}) corresponding
to fixed values of $a_{\alpha}, \, N_{\alpha}$:
\begin{equation}
{\bf V}(Q) = {\bf D}^{-1}{\bf Q}{\bf Z}_0{\bf \Lambda} \left({\bf D}^{-1}{\bf
Q}{\bf Z}_0\right)^*-{\bf h}_0. \label{VQ}
\end{equation}
Here, ${\bf \Lambda}$ and ${\bf Z}_0$ are fixed diagonal and orthogonal ${\cal
N}\times{\cal N}$-matrices respectively; ${\bf Q}$ is arbitrary orthogonal
matrix of the form
\bigskip
\begin{equation}
{\bf Q} = \left(
             \begin{array}{lccr}
               { \begin{picture}(40,40) \put(0,0){\line(1,0){40}}
               \put(0,40){\line(1,0){40}} \put(0,0){\line(0,1){40}}
               \put(40,0){\line(0,1){40}} \put(20,17.5){$ {\bf I} $}
                \end{picture}}&
               { \begin{picture}(5,40) \put(-5,50){\small $(N_1\mbox{-th
               column})$} \put(0,0){$ 0 $} \put(0,15){$ \vdots $}
               \put(0,32.5){$ 0 $}
                \end{picture}}&
               { \begin{picture}(80,40) \put(0,0){\line(1,0){80}}
               \put(0,40){\line(1,0){80}} \put(0,0){\line(0,1){40}}
               \put(80,0){\line(0,1){40}} \put(35,17.5){$ {\bf II} $}
                \end{picture}} &
               { \begin{picture}(5,40) \put(-5,50){\small $({\cal N}\mbox{-th
               column})$} \put(0,15){$ \vdots $} \put(0,32.5){$ 0 $}
                \end{picture}} \\
               { \begin{picture}(40,5) \put(185,0){\small $(N_1\mbox{-th
               row})$} \put(0,0){$ 0 $} \put(15,0){$ \ldots $} \put(37.5,0){$
               0 $}
                \end{picture}}&
               { \begin{picture}(5,5) \put(0,0){$ 1 $}
                \end{picture}}&
               { \begin{picture}(80,5) \put(0,0){$ 0 $} \put(30,0){$ \ldots $}
                \end{picture}}&
               { \begin{picture}(5,5) \put(0,0){$ 0 $}
                \end{picture}} \\
               { \begin{picture}(40,80) \put(0,0){\line(1,0){40}}
               \put(0,80){\line(1,0){40}} \put(0,0){\line(0,1){80}}
               \put(40,0){\line(0,1){80}} \put(15,35){$ {\bf III} $}
                \end{picture}}&
               { \begin{picture}(5,80) \put(0,40){$ \vdots $} \put(0,70){$0$}
                \end{picture}}&
               { \begin{picture}(80,80) \put(0,0){\line(1,0){80}}
               \put(0,80){\line(1,0){80}} \put(0,0){\line(0,1){80}}
               \put(80,0){\line(0,1){80}} \put(35,35){$ {\bf IV} $}
                \end{picture}} &
               { \begin{picture}(5,60) \put(0,0){$ 0 $} \put(0,40){$ \vdots $}
                \end{picture}} \\
               { \begin{picture}(40,5) \put(185,0){\small $({\cal N}\mbox{-th
               row})$} \put(0,0){$ 0 $} \put(15,0){$ \ldots $}
                \end{picture}}&
               { \begin{picture}(5,5) \put(0,0){$ 0 $}
                \end{picture}}&
               { \begin{picture}(80,5) \put(35,0){$ \ldots $} \put(75,0){$ 0 $}
                \end{picture}}&
               { \begin{picture}(5,5) \put(0,0){$ 1 $}
                \end{picture}} \\
             \end{array}
       \right), \label{Qmtx1}
\end{equation}
${\bf h}_0$ is the block-diagonal matrix of the free Hamiltonian involving
submatrices ${\bf h}_0^{(\alpha)}$ (\ref{h0ij}). Thus, the eigenvalues $\left\{
\lambda_j \right\}$ and rows $\left\{ Z_{N_1, \, j} \right\}$,
$\left\{Z_{{\cal N}, \, j} \right\}$ of the eigenvectors matrix ${\bf Z}$ of
the truncated Hamiltonian matrix ${\bf h}'$ specify the potential
(\ref{GPot}), (\ref{PPot}) ${\bf V}$ to within the orthogonal matrix ${\bf Z}$
construction method from given $N_1$-th and ${\cal N}$-th rows. Once this
method has been fixed (e.g. \cite{TMP2}), and the corresponding matrix has
been denoted by ${\bf Z}_0$, the uncertainty of sought-for potential can be
concentrated in the arbitrary orthogonal matrix ${\bf Q}$ (\ref{Qmtx1}).

\par It should be noted that from (\ref{Pab}) it also follows that
the ${\cal P}$-matrix is invariant under exchange of $Z_{N_1, \, j }$ and
$Z_{{\cal N}, \, j }$ signs.

\subsubsection*{Taking into account the bound state wave function}

\par The components $u_{\alpha}$ of the bound state wave function
are represented in the form of expansion
\begin{equation}
u_{\alpha}(\mbox{i}\kappa, \, r) = \sum \limits_{n=0}^{\infty}
b_n^{(\alpha)}(\mbox{i}\kappa) \, \left|\phi_n^{(\alpha)} \right>, \qquad
\alpha =1, \, 2, \label{wfbss}
\end{equation}
where the coefficients $b_n^{(\alpha)}$ for $n_{\alpha} \ge N_{\alpha}-1$ are
given by \cite{BR}:
\begin{equation}
  b^{(\alpha)}_n(\mbox{i}\kappa)= {\cal S}_{\alpha} \,
  C_{n}^{(\alpha)(+)}(\mbox{i} \kappa). \label{bbound}
\end{equation}
Notice that the parameters $\left\{ \lambda_j, \, Z_{N_1, \, j }, \, Z_{{\cal
N}, \, j}, \; j=\overline{1, \, {\cal N}} \right\}$ determine the asymptotic
properties of the bound state wave function. Really, from the equation
\cite{TMP3}
\begin{equation}
\left[ {\bf C}_{N-1}^{(+)} - {\bf {\cal P}}{\bf J}{\bf C}_N^{(+)} \right] \,
{\bf {\cal S}} = 0,
\end{equation}
it follows that
\begin{equation}
\widetilde{\eta} \equiv {\cal S}_2/{\cal S}_1=\frac{\displaystyle
C_{N_1-1}^{(1)(+)} (\mbox{i}\kappa)-{\cal P}_{11}(\varepsilon) \, J_{N_1-1, \,
N_1}^{(1)}(\mbox{i}\kappa) \, C_{N_1}^{(1)(+)}(\mbox{i}\kappa) }
{\displaystyle {\cal P}_{12}(\varepsilon) \, J_{N_2-1, \,
N_2}^{(2)}(\mbox{i}\kappa) \, C_{N_2}^{(2)(+)}(\mbox{i}\kappa) }. \label{eta}
\end{equation}
To derive the expression for ${\cal S}_1$, let us define two auxiliary ${\cal
N}$-component vectors:
\begin{equation}
\begin{array}{c}
x_j = \frac{\displaystyle b_{j-1}^{(1)}(\mbox{i} \kappa) -
b_{j}^{(1)}(\mbox{i}\kappa) } {\displaystyle {\cal S}_1 \,d_{j-1}^{(1)} }, \;
j=\overline{1, \, N_1-1}, \; x_{N_1} = \frac{\displaystyle
C_{N_1-1}^{(1)(+)}(\mbox{i} \kappa)}
{\displaystyle d_{N_1-1}^{(1)} }, \\[5mm]
x_j = \frac{\displaystyle b_{j-N_1-1}^{(2)}(\mbox{i} \kappa) -
b_{j-N_1}^{(2)}(\mbox{i} \kappa) } {\displaystyle {\cal S}_1
\,d_{j-N_1-1}^{(2)} }, \; j=\overline{N_1+1, \, {\cal N}-1}, \; x_{\cal N}
=\frac {\displaystyle \widetilde{\eta} \, C_{N_2-1}^{(2)(+)}(\mbox{i} \kappa)}
{\displaystyle d_{N_2-1}^{(2)} },\\
 \end{array}
\label{x}
\end{equation}
and
\begin{equation}
 \begin{array}{l}
g_j= -\frac{\displaystyle 1} {\displaystyle \kappa^2 + \lambda_j} \left\{
Z_{N_1, \, j} \, d_{N_1-1}^{(1)} \, J_{N_1-1, \, N_1}^{(1)}(\mbox{i}\kappa) \,
C_{N_1}^{(1)(+)}(\mbox{i}\kappa) + \right. \\[2mm]
\phantom{CCCCCCCCCCCCCCC} \left. + \widetilde{\eta} \, Z_{{\cal N}, \, j} \,
d_{N_2-1}^{(2)} \, J_{N_2-1, \, N_2}^{(2)}(\mbox{i}\kappa) \,
C_{N_2}^{(2)(+)}(\mbox{i}\kappa) \right\},
 \; j=\overline{1, \, {\cal N}}.\\
 \end{array}
\label{g}
\end{equation}
With the help of (\ref{x}), (\ref{g}), we can rewrite (\ref{bsol}),
(\ref{bbound}) as
\begin{equation}
{\bf x} = {\bf Z \, g} \label{eqZ}.
\end{equation}
Since ${\bf Z}$ is the orthogonal matrix, the vector ${\bf x}$ norm is equal
to that of ${\bf g}$:
\begin{equation}
{\bf x}^* \, {\bf x}= {\bf g}^* \, {\bf g}. \label{xg}
\end{equation}
Notice that the functions $u_{\alpha}$ (\ref{wfbss}) can be rewritten in the
form:
\begin{equation}
 \begin{array}{c}
  u_1(\mbox{i}\kappa, r)= {\cal S}_1 \left\{ \sum \limits _{n=0} ^{N_1-2}
  x_{n+1} \left| \psi_n^{(1)}\right> +
    \widetilde{u}_1(\mbox{i}\kappa, r) \right\},\\[4mm]
  u_2(\mbox{i}\kappa, r)= {\cal S}_1 \left\{ \sum \limits _{n=0} ^{N_2-2}
  x_{N_1+n+1} \left| \psi_n^{(2)}\right> +
    \widetilde{\eta} \,\widetilde{u}_2(\mbox{i}\kappa, r) \right\},\\
 \end{array}
\label{u1}
\end{equation}
where
\begin{equation}
 \begin{array}{c}
  \widetilde{u}_{\alpha}(\mbox{i}\kappa, r)= u_{\alpha}^{(0)}(\mbox{i}\kappa,
  r)- \sum \limits _{n=0} ^{N_{\alpha}-2} \frac{\displaystyle
  C_n^{(\alpha)(+)}(\mbox{i}\kappa)- C_{n+1}^{(\alpha)(+)}(\mbox{i}\kappa)}
  {\displaystyle d_n^{(\alpha)}}
        \left| \psi_n^{(\alpha)}\right>,\\[4mm]
  u_{\alpha}^{(0)}(\mbox{i}\kappa, r)=\sum \limits _{n=0} ^{\infty}
  C_n^{(\alpha)(+)}(\mbox{i}\kappa) \left| \phi_n^{(\alpha)}\right>= \sum
  \limits _{n=0} ^{\infty} \frac{\displaystyle
  C_n^{(\alpha)(+)}(\mbox{i}\kappa)- C_{n+1}^{(\alpha)(+)}(\mbox{i}\kappa)}
  {\displaystyle d_n^{(\alpha)}}
        \left| \psi_n^{(\alpha)}\right>.\\
 \end{array}
\label{u2}
\end{equation}
Substitution of Eq. (\ref{xg}-\ref{u2}) into the unit norm of the wave
function condition
\begin{equation}
\int \limits_0^{\infty} \left| u_1(\mbox{i}\kappa, \, r) \right|^2 dr + \int
\limits_0^{\infty} \left| u_2(\mbox{i}\kappa, \, r) \right|^2 dr =1
\end{equation}
gives:
\begin{equation}
 \begin{array}{l}
{\cal S}_1^{-2}= {\bf g}^* \, {\bf g}- \left|\frac{\displaystyle
C_{N_1-1}^{(1)(+)}(\mbox{i} \kappa)} {\displaystyle d_{N_1-1}^{(1)} }
\right|^2 - \left|\frac{\displaystyle \widetilde{\eta}
C_{N_2-1}^{(2)(+)}(\mbox{i} \kappa)}
{\displaystyle d_{N_2-1}^{(2)} } \right|^2 + \\[5mm]
\phantom{CCCCCCCCCCCCCCCCCCCC} \int \limits_0^{\infty} \left|
\widetilde{u}_1(\mbox{i}\kappa, \, r) \right|^2 dr + \int \limits_0^{\infty}
\left|\widetilde{\eta}
\widetilde{u}_2(\mbox{i}\kappa, \, r) \right|^2 dr.\\
 \end{array}
\label{eS1}
\end{equation}

\par In the framework of our method the phase-equivalent transformation
(\ref{VQ}), (\ref{Qmtx1}) is a tool for the influence on the potential
(\ref{GPot}), (\ref{PPot}) off-shell behaviour. If we only be interested in
describing the bound state wave function properties, we may vary the first
$N_{\alpha}-1$ coefficients of the expansions (\ref{wfbss}) in view of the
condition (\ref{xg}). For ${\cal N} \ge 4$, the resulting set $\left\{
b_n^{(\alpha)}, \; n=\overline{0, \, N_{\alpha}-2} \right\}$ does not uniquely
specify (see (\ref{eqZ})) the matrix ${\bf Q}$ (\ref{Qmtx1}), since its
unknown submatrix $\overline{\bf Q}$
\begin{equation}
\overline{\bf Q} = \left(
             \begin{array}{c}
                \begin{picture}(120,120)
                \put(0,0){\line(1,0){120}} \put(0,80){\line(1,0){120}}
                \put(0,120){\line(1,0){120}} \put(0,0){\line(0,1){120}}
                \put(120,0){\line(0,1){120}} \put(40,0){\line(0,1){120}}
                \put(19,100){$ {\bf I} $} \put(75,100){$ {\bf II} $}
                \put(15,40){$ {\bf III} $} \put(75,40){$ {\bf IV} $}
                \end{picture}\\
            \end{array}
       \right) \label{Qmat_}
\bigskip
\end{equation}
is of order ${\cal N}-2$.

\section{Results}

\par As an application, we invert the $NN$ elastic scattering
data.  A set of Nijmegen \cite{Nijmegen} phase shifts in the $0 - 350$~MeV
energy range is taken as input to calculate the potentials. The obtained $pp$
potentials parameters in the channels $^1S_0$ and $^3P_2-^3F_2$ are listed in
Table~I. In Fig.~1 we show the phase shifts resulting from the Nijmegen
partial wave analysis \cite{Nijmegen} (crosses) and the phase shifts obtained
from our potentials (solid line).

\par In the $^1 S_0$ channel for a chosen $a$ and $N$ the pair of unknown
``external'' parameters $\lambda_4, \, Z_{4, \, 4}$ has been obtained. Since
the $Z_{4, \, 4}$ value is fixed by the normalization of the vector $\left\{
Z_{4, \, j}\right\}$ condition, only free parameter $\lambda_4$ remains.

\par In the channel $^3P_2-^3F_2$ the absolute values of the phase shift
$\overline{\delta}_3$ are very small. For the relative deviations of the
$\overline{\delta}_3$ values from ones of \cite{Nijmegen} to be comparable to
those for the phase shift $\overline{\delta}_1$, the value of $N_2$ must be
chosen sufficiently large (see Table~I). For a chosen combination of the
$a_{\alpha}, \, Na_{\alpha}$, we obtain three triplets of the ``external''
parameters $\left\{ \lambda_j, \, Z_{2, \, j}, \, Z_{8, \, j}, \; j=6, \, 7,
\, 8\right\}$. The values of $\lambda_6$ and $\lambda_7$ are obtained by
making it equal to $\lambda_2^{(1)}$ and $\lambda_5^{(2)}$ calculated in the
potentials construction in the uncoupled channels $^3P_2$ and $^3F_2$
respectively. The parameters $\lambda_8, \, Z_{2, \, 8}, \, Z_{8, \, 8}$
values were calculated in a similar way to those of the ``internal''; to do
this the phase shifts are extrapolated with functions which decay
asymptotically. Finally, the orthonormalization conditions (\ref{Norm}) allow
the free parameters number to be reduced to one. For the latter we used $Z_{2,
\, 7}$.

\par The obtained values for the $np$ potential parameters in the channel
$^3S_1-^3D_1$ are given in Table II. The phase shifts and the mixing parameter
corresponding to our potential are represented by solid lines in Fig.~2.

\par Here, we also have obtained three triplets $\left\{ \lambda_j \right.$,
$\left. Z_{2, \, j}, \, Z_{8, \, j}\right.$, $\left. j=6, \, 7, \, 8\right\}$
of the ``external'' parameters. The values $\lambda_6$ and $\lambda_7$ are
equal to $\lambda_4^{(2)}$ and $\lambda_4^{(1)}$ obtained in the uncoupled
channels $^3D_1$ and $^3S_1$ respectively. Next, orthonormalization conditions
(\ref{Norm}) have reduced the unknown parameters number to four, three of
which $Z_{4, \, 7}, \, Z_{4, \, 8}, \, Z_{8, \, 8}$ are variable. Further, the
combination of $Z_{4, \, 8}, \, Z_{8, \, 8}$ has an effect on the phase shift
$\overline{\delta}_0$ and mixing parameter $\overline{\varepsilon}_1$
behaviour in the neighborhood of $\epsilon=0$, whereas $Z_{4, \, 7}$ is
responsible for $\overline{\varepsilon}_1$ curve form at the right of the
interval $[0, \, \epsilon_0]$. Then the parameter $\lambda_8$ fits Eq.
(\ref{Bcond}) with the deuteron binding energy $\varepsilon = - \kappa^2 =
-.0536401 \, fm^{-2}$.

\par The deuteron wave function of the potential Nijm II \cite{Nijmegen} is
employed as input in the $np$ potential constructing. It is seen from
(\ref{u1}), (\ref{u2}) and the relation $\mathop{u^{(0)}_S(\mbox{i}\kappa,\,
r)} \limits_{r \to \infty}~ \longrightarrow~ e^{-\kappa \, r}$ that the
asymptotic $S$-state normalization $A_S$ is equal to $S_1$, i. e. depends on
the module of the vector ${\bf g}$ (\ref{g}) (see (\ref{eS1})). The last
component $g_8$ containing the parameters $Z_{4, \, 8}, \, Z_{8, \, 8}, \,
\lambda_8$ makes a major contribution to the vector ${\bf g}$ norm (see
Table~III). $A_S$ is found to be the most sensitive to the change in $Z_{4, \,
8}$, whereas the $D/S$-state ratio $\eta$ ($\eta=-\widetilde{\eta}$
(\ref{eta}), since $\mathop{u^{(0)}_D (\mbox{i}\kappa, \, r)} \limits_{r \to
\infty}~ \longrightarrow~ -e^{-\kappa \, r}$) is strongly dependent on the
value of $Z_{8, \, 8}$.

\par In order to get the right behaviour of the wave function in the
interior domain we use the phase-equivalent transformation (\ref{VQ}),
(\ref{Qmtx1}). In so doing the vector ${\bf x}$ (\ref{x}) components $\left\{
x_1, \, x_2, \, x_3\right\}$ and $\left\{ x_5, \, x_6, \, x_7 \right\}$ (which
within the factor ${\cal S}_1$ are equal to the first three coefficients of
the $S$ and $D$ waves expansions (\ref{wfbss}) on the basis orthonormal
functions (\ref{bf1}) ) are varied in view of Eq. (\ref{xg}). In Fig.~3 the
deuteron wave function $S$ and $D$ components (functions $u_S$ and $u_D$
respectively) of our potential with the parameters in Tables II, III are drawn
(dotted line). There we compare them with the wave functions of the Nijm II
potential \cite{Nijmegen} (solid line). The deuteron parameters calculated
with our wave function and those from the Nijm II potential are listed in
Table~IV.

\section{Summary}

\par Approximate inverse scattering method  within
$J$-matrix approach \cite{TMP1,TMP2} has been extended to the case of the
Laguerre form factors. As a result the method has become applicable to the
inversion of charge particles scattering. The ${\cal P}$-matrix has a dominant
role in our procedure. Similar to the $R$-matrix, the ${\cal P}$-matrix
contains requisite information for the potential constructing and on the other
hand is uniquely determined by the $S$-matrix. Besides, the ${\cal P}$-matrix
determines the asymptotic normalization constants of the bound state wave
functions.

\par In this work the phase-equivalent transformation \cite{PHT,TMP2} has been
modified to be valid for separable potentials with the Laguerre form factors.
The phase-equivalent transformation was used for the description of the
deuteron wave function of the realistic Nijm II \cite{Nijmegen} nucleon-nucleon
potential . In doing this, we were interested in the behaviour of the wave
function in the internal domain rather than in fitting the values of the
quadrupole moment ${\cal Q}_d$ or the $D$-state probability $p_D$. The
discrepancy between the deuteron wave functions of our potential and that of
Nijm II potential can be removed by increasing of desired potential
(\ref{GPot}), (\ref{PPot}) rank ${\cal N}$.

\subsection*{Acknowledgment}
\par It is pleasure to thank Prof. A~.M.~Shirokov and Prof. A.~I.~Mazur for
helpful discussions. This work was supported in part by the State Program
``Universities of Russia'', project No 992306.

\newpage

\begin{center}
{\Large Tables}
\end{center}

Table I. $pp$ potentials parameters values in the channels $^1S_0$ and
$^3P_2-^3F_2$.
$$
 \begin{array}{ccc|ccc}
\hline \hline
& & & &  &\\
\phantom{ } & \multicolumn{2}{c}{l=0, \: a=2.5 fm^{-1}, \: N=4} &
 \multicolumn{3}{|c}{ \begin{array}{c} l_1=1, \: a_1=5 fm^{-1}, \: N_1=2, \\
  l_2=3, \: a_2 =3 fm^{-1}, \: N_2=6\\
 \end{array}
}\\
  & & & & & \\
\hline
  & & & & & \\
j & Z_{N, \, j} & \lambda_j &
Z_{N_1, \, j} & Z_{ {\cal N}, \, j} & \lambda_j\\
 & &  (fm^{-2}) & & &  (fm^{-2})\\
\hline
  \begin{array}{c}
   1\\ 2\\ 3\\ 4\\ 5\\ 6\\ 7\\ 8\\
  \end{array}
 & \begin{array}{l}
.1558523299\\ .4062499094\\ .6517820375\\ .6211692507\\ \\ \\ \\ \\
   \end{array}
 & \begin{array}{l}
.08517568687\\ .6366888556\\  2.999305831\\ 22.\\ \\ \\ \\ \\
   \end{array}
 & \begin{array}{l}
.002378372193\\ .006334410556\\ .02352440740\\ .4982479821\\ .02790963713\\
.01090380937\\ .8657366805\\ .02742131063\\
   \end{array}
 & \begin{array}{l}
-.2021238864\\ \phantom{-}  .2927989544\\ -.3709560916\\
\phantom{-} .03288283700\\ -.4402591159\\ \phantom{-} .4980705527\\
\phantom{-} .01462158020\\ -.5409333126\\
   \end{array}
 & \begin{array}{l}
.3171350908\\ .7864286930\\ 1.607103324\\ 2.259459783\\ 3.146276272\\
6.48\\ 13.25\\ 16.20600628\\
  \end{array}\\
\hline \hline
\end{array}
$$

\vspace{5mm}

Table II. $np$ potential parameters values in the channel $^3S_1-^3D_1$.
$$
\begin{array}{cccc}
\hline \hline
 & & &\\
\multicolumn{4}{c}{
 \begin{array}{c}
  l_1=0, \: a_1=2.8 fm^{-1}, \: N_1=4,\\
  l_2=2, \: a_2 =3 fm^{-1}, \: N_2=4\\
 \end{array}
}\\
 & & &\\
\hline
 & & & \\
j & Z_{N_1, \, j} & Z_{ {\cal N}, \, j} & \lambda_j\\
 & & &  (fm^{-2})\\
\hline
  \begin{array}{c}
 1\\ 2\\ 3\\ 4\\ 5\\ 6\\ 7\\  8\\
  \end{array}
 & \begin{array}{l}
.02017320192\\ .3982489552\\ .01890700216\\ .6535995282\\ .1314698511\\
.08780677309\\ .6152722974\\ .09943159709\\
   \end{array}
 &\begin{array}{l}
-.2989525605\\ \phantom{-} .02212234580\\ -.4605017812\\
\phantom{-} .09501129016\\ -.5784348737\\ \phantom{-} .5929301489\\
-.05339635194\\ \phantom{-} .006685636351\\
   \end{array}
 & \begin{array}{l}
\phantom{-} .4678515858\\ \phantom{-} .6496845646\\  \phantom{-} 1.501863086\\
\phantom{-} 3.492094508\\ \phantom{-} 4.238758885\\
\phantom{-} 12.75\\  \phantom{-} 35.\\ -.03571632149\\
  \end{array}\\
\hline \hline
\end{array}
$$

\newpage
Table III. The components of the vectors ${\bf g}$ (\ref{g}) and ${\bf x}$
(\ref{x}) in $fm^{1/2}$.
$$
\begin{array}{ccc}
\hline \hline
  & &\\
j & g_j & x_j\\
 & & \\
\hline
  \begin{array}{c}
 1\\ 2\\ 3\\ 4\\ 5\\ 6\\ 7\\  8\\
  \end{array}
 &\begin{array}{l}
-.002345654140\\ -.1089332559\\ \phantom{-} .0002896455277\\ -.03560896022\\
-.004683742390\\ -.001726119865\\ -.003354599822\\ -1.067801341\\
\end{array}
 & \begin{array}{l}
\phantom{-} .6665\\ -.69\\ \phantom{-} .341506600\\
-.1757029190\\ \phantom{-} .265\\ \phantom{-} .025\\
\phantom{-} .045\\ -.01049929153\\
  \end{array}\\
\hline \hline
\end{array}
$$

\vspace{5mm}

Table IV. Deuteron properties: $D/S$-ratio $\eta$, asymptotic $S$-state
normalization $A_S$ in \\$fm^{-1/2}$, $D$-state probability $p_D$ in $\%$, and
quadrupole moment ${\cal Q}_d$ in $fm^2$.
$$
 \begin{array}{ccccc}
\hline \hline
& & & & \\
& \eta & A_S & p_D & {\cal Q}_d\\
& & & &  \\
\hline
\mbox{(this article)} & .0252 & .8845 & 5.729 & .28185 \\
& & & &  \\
\mbox{Nijm II} & .0252 & .8845 & 5.635 & .2707 \\
\hline \hline
\end{array}
$$

\newpage

\begin{center}
{\Large Figure captions}
\end{center}

Fig.~1. $pp$ phase shifts in the channels $^1S_0$ and $^3P_2-^3F_2$ from the
Nijmegen \cite{Nijmegen} partial-wave analysis (crosses) and the potential
(\ref{GPot}), (\ref{PPot}) with the parameters listed in Table~I (solid line).

\vspace{5mm}

Fig.~2. $np$ phase shifts in the channel $^3S_1-^3D_1$ from the Nijmegen
\cite{Nijmegen} partial-wave analysis (crosses) and the potential
(\ref{GPot}), (\ref{PPot}) with the parameters listed in Table~II (solid line).

\vspace{5mm}

Fig.~3. The $S$ and $D$ components of the deuteron wave functions of the
Nijm~II potential \cite{Nijmegen} (solid line) and of the potential
(\ref{GPot}), (\ref{PPot}) with the parameters listed in Tables~II, III
(dotted line).

\newpage
\includegraphics[bbllx=110,bblly=450,scale=1.125]{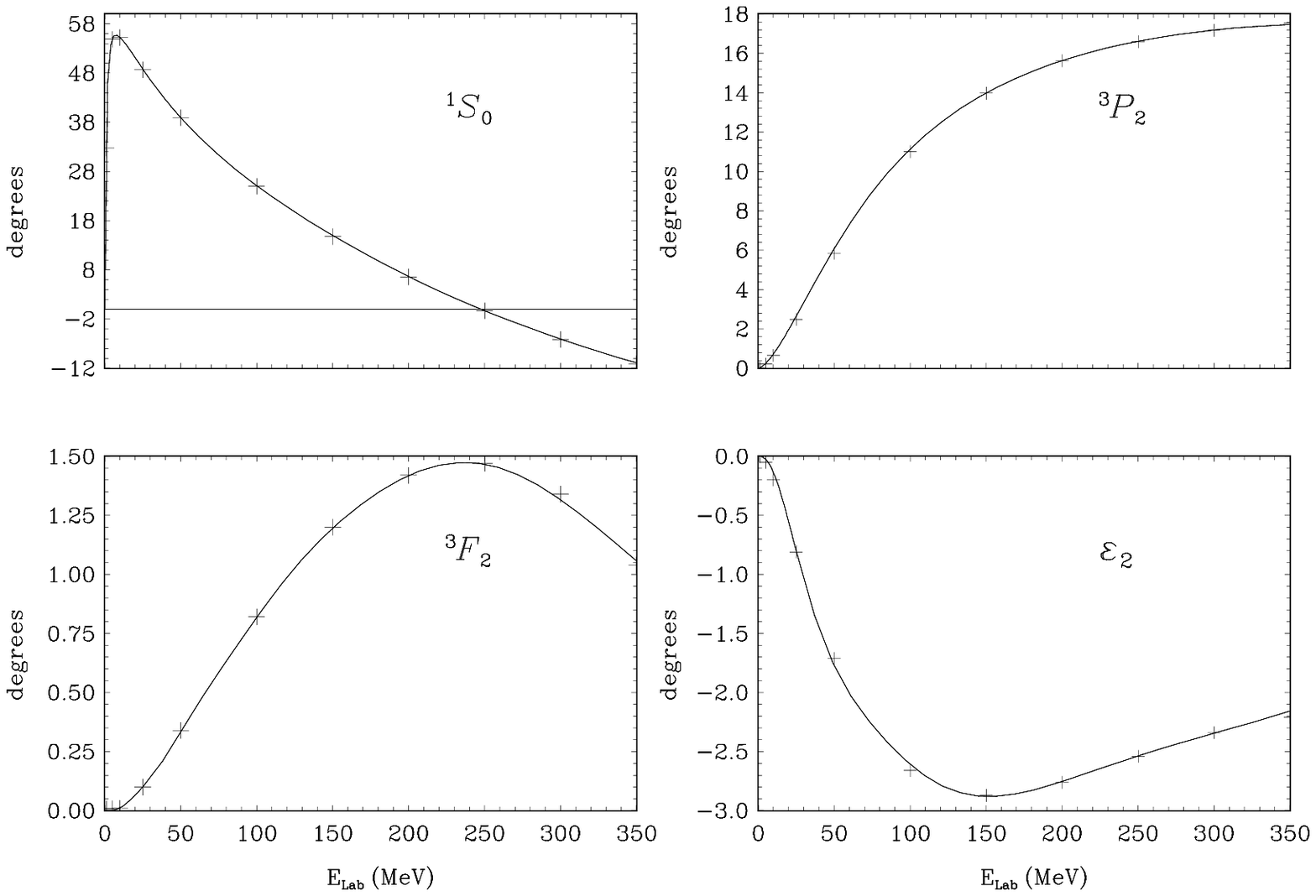}
\vfill
\begin{center}
{\Large \bf Figure 1.}
\end{center}

\newpage
\includegraphics[bbllx=54,bblly=385,scale=2.2]{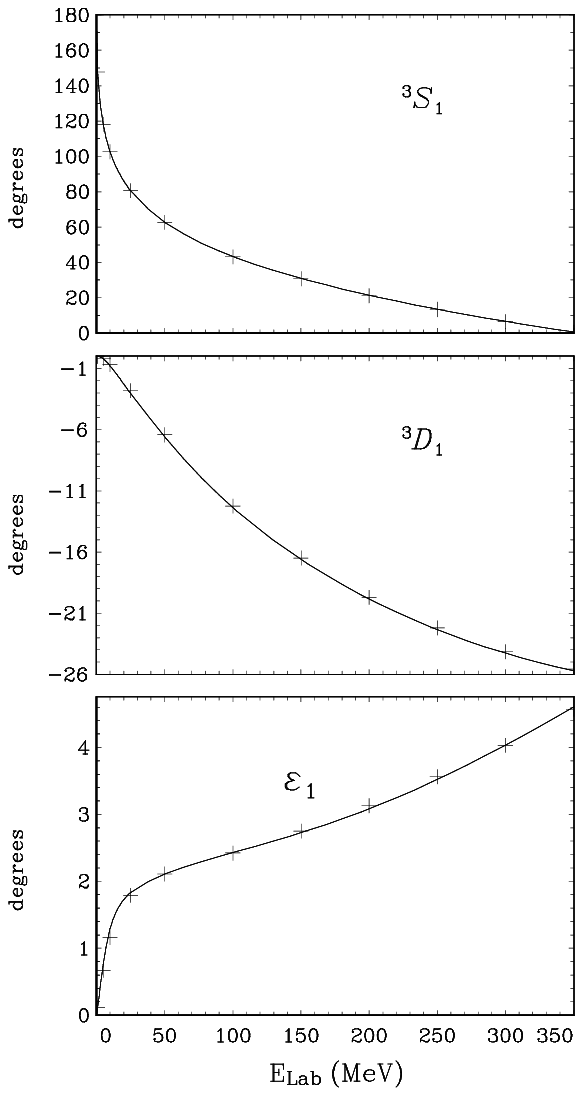}
\vfill
\begin{center}
{\Large \bf Figure 2.}
\end{center}

\newpage
\includegraphics[bbllx=55,bblly=450,scale=1.65]{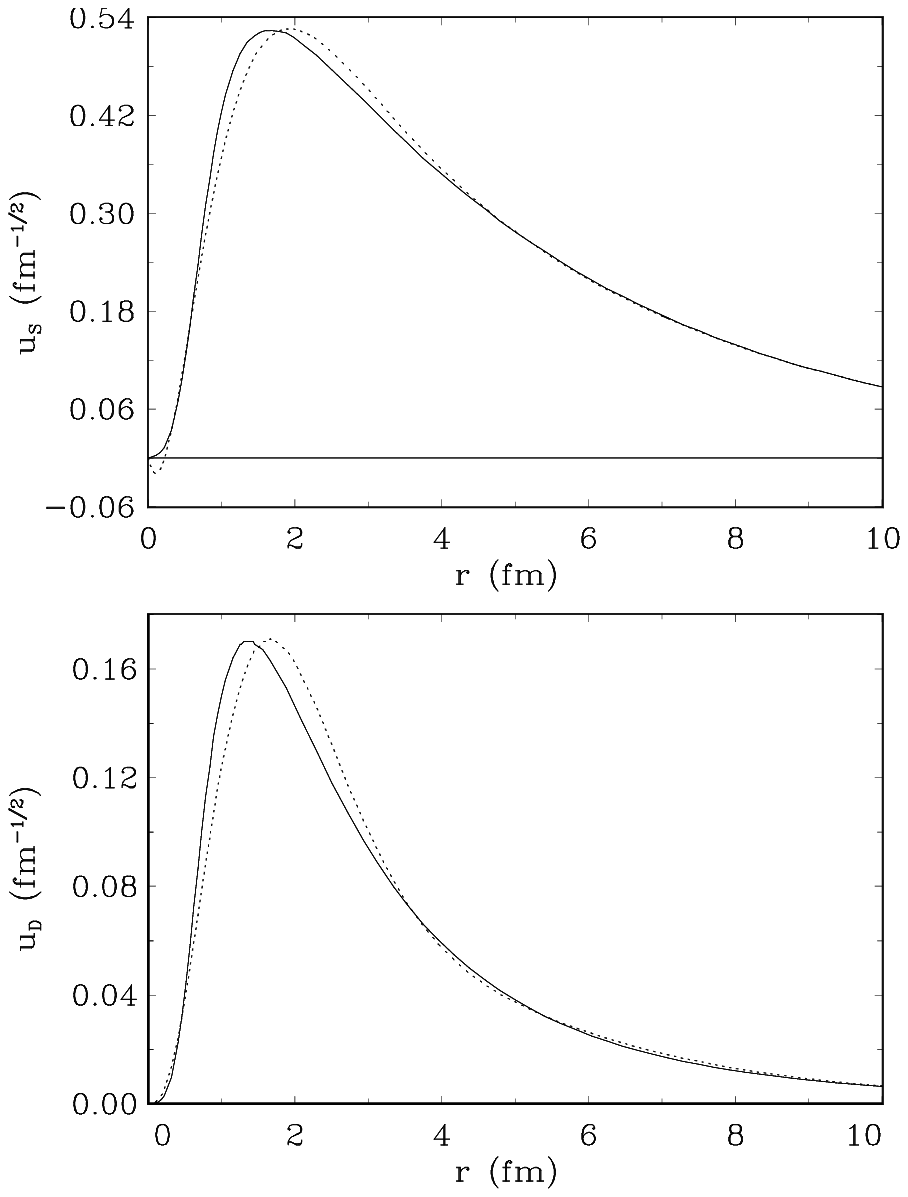}
 \vfill
\begin{center}
{\Large \bf Figure 3.}
\end{center}


\begin{thebibliography}{99}

\bibitem{SepNN}
 N.~H.~Kwong, H.~S.~K\"{o}hler, Phys. Rev. C {\bf 55}, 1650 (1997).

\bibitem{Shell}
 P.~Navratil, B.~R.~Barrett, Phys.Rev. C {\bf 54} 2986 (1996).

\bibitem{Nijmegen}
 V.~G.~J.~Stoks, R.~A.~M.~Klomp, C.~P.~F.~Terheggen, and J.~J. de Swart, Phys.
 Rev. C, {\bf 49}, 2950 (1994).

\bibitem{TB}
 Z.~Papp and W.~Plessas, Phys. Rev. C {\bf 54}, 50 (1996).

\bibitem{TMP1}
 S.~A.~Zaitsev, Teoret, Mat. Fiz. {\bf 115}, 263 (1998) [Theor. Math. Phys.
 {\bf 115}, 575 (1998)].

\bibitem{TMP2}
 S.~A.~Zaitsev, Teoret, Mat. Fiz. {\bf 121}, 424 (1999) [Theor. Math. Phys.
 {\bf 121}, 1617 (1999)].

\bibitem{Jmtx}
 H.~A.~Yamani, L.~Fishman, J. Math. Phys. {\bf 16}, 410 (1975).

\bibitem{AVMRG}
 G.~F.~Filippov, I.~P.~Okhrimenko, Yad. Fiz. {\bf 32}, 932 (1980) [Sov. J.
 Nucl. Phys., {\bf 32}, 480 (1980)].

\bibitem{HOR}
 Yu.~I.~Nechaev and Yu.~F.~Smirnov, Yad. Fiz. {\bf 35}, 1385 (1982) [Sov. J.
 Nucl. Phys., {\bf 35}, 808 (1982)].

 \bibitem{PSE}
 J.~R\'{e}vai, M.~Sotona, J.~Zofka, J.Phys. G {\bf 11}, 745 (1985).

\bibitem{Papp}
 Z.~Papp, J. Phys. A {\bf 20}, 153 (1987); Phys. Rev. C {\bf 38}, 2457 (1988).

\bibitem{BR}
 J.~T.~Broad, W.~P.~Reinhardt,  J.Phys. B {\bf 9}, 1491 1976.

\bibitem{Zakhariev}
 B.~N.~Zakhariev, A.~A.~Suzko, Direct and inverse problems. In: Potentials in
 quantum scattering. 2-nd ed. Berlin, Heidelberg, New York: Springer-Verlag,
 1990.

\bibitem{PHT}
 Yu.~A.~Lurie and A.~M.~Shirokov, Izv. RAN, Ser. Fiz. {\bf 61}, 2121 (1997)
 [Bul. Rus. Acad. Sci., Phys. Ser. {\bf 61}, 1665 (1997)].

\bibitem{Broad}
 J.~T.~Broad, Phys. Rev. A {\bf 31}, 1494 (1985).

\bibitem{Rmtx}
 A.~M.~Lane, A.~M.~Thomas, Rev. Mod. Phys. {\bf 30}, 257 (1958).

\bibitem{Sk}
 V.~A.~Skorobogatko, The Theory of Branching Continued Fractions and its
 Applications to Computional Mathematics [in Russian], Moscow (1983).

\bibitem{TMP3}
A.~M.~Shirokov, Y.~F.~Smirnov, S.~A.~Zaytsev, Teoret, Mat. Fiz. {\bf 117}, 227
(1998) [Theor. Math. Phys. {\bf 117}, 1291 (1998)].

\end{thebibliography}
\end{document}